\documentstyle[12pt,epsfig]{article}

\newcommand{\be}{\begin{equation}}
\newcommand{\bea}{\begin{eqnarray}}
\newcommand{\eea}{\end{eqnarray}}
\newcommand{\ba}{\begin{array}}
\newcommand{\ea}{\end{array}}
\newcommand{\ee}{\end{equation}}

\newcommand{\cN}{{\cal N}}

\expandafter\ifx\csname mathbbm\endcsname\relax

\else Kim

\fi
\def\appendix{{\newpage\section*{Appendix}}\let\appendix\section%
        {\setcounter{section}{0}
        \gdef\thesection{\Alph{section}}}\section}

\begin{document}

\begin{titlepage}
\hfill
\vbox{
    \halign{#\hfil         \cr
           CERN-TH/2001-166 \cr
           UNIL-IPT-01-09  \cr
           hep-th/0106255  \cr
           } 
      }  
\vspace*{20mm}
\begin{center}
{\Large {\bf  Supergravity, Non-Conformal Field Theories and Brane-Worlds}\\} 
\vspace*{15mm}
\vspace*{1mm}
Tony Gherghetta$^a$  and Yaron Oz$^b$ \\
\vspace*{1cm} 
{\it {$^{a}$Institute of Theoretical Physics, University of Lausanne, \\
CH-1015 Lausanne, Switzerland}}\\
\vspace*{5mm}
{\it {$^{b}$Theory Division, CERN \\
CH-1211 Geneva  23, Switzerland}}\\

\vspace*{.75cm}
\end{center}

\begin{abstract}

We consider the supergravity dual descriptions of non-conformal 
super Yang-Mills theories realized on the world-volume of 
D$p$-branes. We use the dual description to 
compute stress-energy tensor and current correlators.
We apply the results to the study of dilatonic brane-worlds described 
by  non-conformal field theories coupled to gravity.
We find that brane-worlds based on D4 and D5 branes exhibit a 
localization of gauge and gravitational fields.
We calculate the corrections to the Newton and Coulomb laws in these 
theories. 

\end{abstract}
\vskip 4cm

June 2001

\end{titlepage}

\newpage

\section{Introduction}

The AdS/CFT correspondence relates a conformal field theory in $d$ dimensions
to string theory on $AdS_{d+1}$ space \cite{malda,Gubser:1998bc,Witten:1998qj}
(For a review see \cite{agmoo}).
The radial coordinate of $AdS_{d+1}$, often denoted by $U$, is interpreted as 
an energy scale from the point of view of the $d$-dimensional conformal field theory. 
The canonical example is the $\cN=4$ SCFT in four-dimensions realized
on the worldvolume of D3-branes.
The dual description being 
type-IIB string theory on $AdS_5 \times S^5$.
The worldvolume theory of D$p$-branes when $p\neq 3$ is a non-conformal super Yang-Mills
theory (SYM) with sixteen supercharges. These theories have dual descriptions which vary as we
move in the energy scale $U$ \cite{imsy}.
They have been much less studied compared to the conformal field theories.

Another line of research is the study of brane-world 
scenarios
\cite{rsII,Verlinde:2000fy}. We will consider a brane-world to be a brane
(not necessarily a D$p$-brane) located at $U=U_{*}$.
When $U_{*}$ is infinite and the brane is a  D$p$-brane 
we recover the above, where the D$p$-brane theory is a Yang-Mills
field theory without gravity. When $U_{*}$ is finite, the theory on the brane is
a Yang-Mills theory coupled to gravity with a UV cutoff $\Lambda = U_{*}$.
As in the duality between gauge theory and gravity, the brane-world 
case with $p=3$ has been extensively studied, as a conformal field theory coupled to gravity
system. 
This brane-world scenario 
exhibits a localization of gravity on the brane-world. However a bulk 
gauge field is not localized in this scenario.

In this paper we will consider brane-worlds based on D$p$-branes when $p\neq 3$. 
We will start by
considering  the supergravity dual descriptions and use it to
compute stress-energy tensor and current correlators of the non-conformal field 
theories.
We will then apply the results to the study of dilatonic brane-worlds described by 
non-conformal
field theories coupled to gravity.
We will find that 
brane-worlds based on D4 and D5 branes exhibit a localization of gauge and gravitational
fields, and we calculate 
the corrections to the Newton and Coulomb laws in these theories.

The paper is organized as follows.
In section 2 
we will consider the worldvolume theory of $N$ coincident
D$p$-branes in type-II string theory 
using the dual supergravity description. We will compute correlation functions of the
stress-energy tensor and global symmetry currents.
In section 3 we will consider 
brane-world scenarios with a brane located at $U=U_{*}$.
When $p\neq 3$ the theory on the brane-world is dual to a $(p+1)$-dimensional
non-conformal super Yang-Mills theory coupled to gravity
with a cutoff $\Lambda = U_{*}$.
These brane-world scenarios are sometimes called dilatonic domain walls.
We will analyse these brane-world scenarios with and without additional
compactification of brane worldvolume coordinates.
We will use the results of the previous section in order to compute the corrections
to Newton and Coulomb laws.
We will also derive these corrections in a slightly different approach and discuss
the localisation of gauge and gravitational fields.
We find that brane-worlds based on D4 and D5 branes exhibit a 
localization of both
gauge and gravitational fields.
Section 4 is devoted to a more detailed discussion on brane-world 
scenarios based on dilatonic domain walls.
Some details of Green's functions computations are outlined in the appendix.

\section{D$p$-branes}
In this section we will consider the worldvolume theory of $N$ coincident
D$p$-branes in type-II string theory 
using the dual supergravity description. We will compute correlation functions of the
stress-energy tensor and global symmetry currents.

\subsection{The dual supergravity description}

Consider a system of 
$N$ coinciding flat D$p$-branes of type-II string theory.
We denote the ten-dimensional space coordinates by $x_0,...,x_9$. 
The $p+1$ worldvolume coordinates are taken to be $x_0,..., x_{p}$.
The field theory limit of this D$p$-brane system is defined by 
taking the string length, $l_s \rightarrow 0$ while keeping 
the Yang-Mills coupling, $g_{YM}^2 = g_s l_s^{p-3}$ fixed, where $g_s$
is the string coupling \cite{imsy}.
For $p < 6$ this is a limit that decouples the open string degrees of freedom
attached to the D$p$-branes worldvolume from the bulk closed string degrees of freedom.
The worldvolume theory is a $(p+1)$-dimensional $U(N)$ super Yang-Mills theory with 
sixteen supercharges.
In the following we will consider the large $N$ t' Hooft limit, where 
we denote the t' Hooft parameter by $\lambda = g_{YM}^2 N$.

This system has a conjectured dual string description \cite{imsy}.
In the supergravity approximation 
the (string-frame) metric describing the system is given by 
\bea
   ds^2 &=& l_s^2 \left(\frac{U^{\frac{1}{2}(7-p)}}
   {\lambda^{\frac{1}{2}}} dx_{||}^2 +
   \frac{\lambda^{\frac{1}{2}}}{U^{\frac{1}{2}(7-p)}} dU^2 
 + \lambda^{\frac{1}{2}}U^{\frac{1}{2}(p-3)}
d \Omega_{8-p}^2 \right) \ , \nonumber\\
e^{\phi} &=& \frac{\lambda^{\frac{1}{4}(7-p)}}{N} U^{\frac{1}{4}(7-p)(p-3)} \ .
\label{metric}
\eea
Note that we have omitted 
numerical factors that will be irrelevant for our discussions later.
The $p+1$ coordinates of the D$p$-brane worldvolume are denoted by $x_{||}$, 
and have 
dimension of length. The radial coordinate $U$ has dimensions of mass, 
and from the
D$p$-branes worldvolume field theory point of view it plays the role
of an energy scale. Large values of $U$ correspond to the UV regime and 
small values correspond to the IR regime of the Yang-Mills theory.
The angular coordinates of the $(8-p)$-dimensional sphere are denoted
by $\Omega_{8-p}$. They are dimensionless in the above notation.
The field $\phi$ is the ten-dimensional dilaton.  

We will be mostly interested in the non-conformal cases $p \neq 3$. 
In these cases the metric (\ref{metric}) 
has the isometry group $ISO(1,p)\times SO(9-p)$.
From the Yang-Mills theory point of view  $ISO(1,p)$ is the Poincare 
symmetry while $ SO(9-p)$ is the R-symmetry group.
The Yang-Mills coupling is dimensionful, $[g_{YM}^2] = [L]^{p-3}$, 
and we can define a dimensionless expansion parameter of the system 
\be
g_{eff}^2 = g_{YM}^2 N U^{p-3}= \lambda U^{p-3}~.
\ee
The Yang-Mills theory is strongly coupled when $g_{eff}> 1$.

The scalar curvature of the metric (\ref{metric}) is given by 
\be
l_s^2{\cal R} \sim \frac{1}{g_{eff}} \ .
\label{curv}
\ee
We see
from (\ref{curv}) that
the curvature expansion in the dimensionless parameter $l_s^2{\cal R}$ on the supergravity
side corresponds to a strong coupling expansion of the gauge theory in the inverse
of the effective Yang-Mills bare coupling.
It is a good expansion when $l_s^2{\cal R}\ll 1$, i.e. in the regime where the Yang-Mills
field theory is strongly coupled.
When $g_{eff}^2 \ll 1$ we have a large curvature. In this regime we cannot trust the supergravity
description. However, the Yang-Mills perturbation theory in $g_{eff}$ is valid in this regime.
The overlap regime is when $g_{eff} \sim 1$.

The  effective string coupling $e^{\phi}$ in (\ref{metric}) can be
expressed in terms of $g_{eff}$ as 
\be
e^{\phi} \sim \frac{g_{eff}^{\frac{1}{2}(7-p)}}{N} \ .
\label{ep}
\ee
We see from (\ref{ep}) that the string 
loop expansion  corresponds to the $1/N$ expansion of the Yang-Mills theory.
This expansion is good when  the effective string coupling
is small  $e^{\phi}\ll1$ .

Thus, 
the regime where the supergravity description of the system
given by (\ref{metric}) is valid is dictated by the conditions
that both the curvature (\ref{curv}) and  effective string coupling (\ref{ep}) are small.
This implies that 
\be
1\ll g_{eff}^2 \ll N^{\frac{4}{7-p}} \ .
\label{regime}
\ee
We will work in this regime.
Moreover, we will consider the cases $p <6$. When $p\geq 6$
the above field theory limit is no longer 
a decoupling limit, and the brane worldvolume theory does
not decouple from the bulk physics. This can be seen, for instance, by the fact that
the eleven-dimensional Planck length $l_{11}^3 = g_s l_s^3$ is fixed in the limit. Therefore,
gravity does not decouple in eleven-dimensions, which provides the proper  description of the system
in UV~\cite{imsy}. Another way is to compute the absorption cross section 
for graviton 
scattering on the branes and see that
it does not vanish in the limit \cite{aio}.
From the brane-world scenario we will see that gravity and 
gauge fields cannot be localized.

\subsection{Stress-energy tensor correlators}

Let us now use the dual supergravity description in order
to compute the two-point function of the stress-energy tensor
of the Yang-Mills field theory on the D$p$-branes worldvolume. 
Unlike the conformal case $p=3$, the two-point function changes as we 
vary the length scale $|x| \equiv U^{-1}$.
The regime of validity of the supergravity computation is given by (\ref{regime}).
As noted above this is a strong coupling regime of the gauge theory.
The results that we will find will match, up to a numerical factor, with
 the perturbative Yang-Mills expectation 
in the crossover regime.

The basic idea of the correspondence between string theory and gauge theory
is the identification, in the supergravity approximation,  of
the generating functional of the connected Green's functions in the 
gauge theory with the minimum 
of the supergravity action subject to certain boundary conditions
\cite{Gubser:1998bc,Witten:1998qj}.
For each operator on the Yang-Mills side we identify a field on the 
supergravity (string) side.
The bulk field that couples to the stress-energy tensor of the gauge theory
is the graviton.  
In the computation of the generating functional for
the stress-energy tensor correlators we need to 
solve the graviton field equation in the background
(\ref{metric}) subject to some boundary conditions and insert it in the 
supergravity action. One decomposes the metric
\begin{equation}
     g_{\mu\nu} = {\bar g}_{\mu\nu} + h_{\mu\nu},~~~~~ \mu,\nu=0,\dots,9
\end{equation}
where ${\bar g}_{\mu\nu}$ denotes the background metric (\ref{metric})
and $h_{\mu\nu}$ denotes the perturbation.
The stress-energy tensor $T_{\mu\nu}$ of the brane worldvolume field theory 
couples to the bulk graviton as
\be
S_{int} \sim \int d^{p+1}x T^{\mu\nu}h_{\mu\nu} \ .
\ee

We can decompose the graviton field as
\begin{equation}
    h_{\mu\nu}(x,U,\Omega) = \epsilon_{\mu\nu} e^{i k\cdot x} h(U) 
    Y(\Omega_{8-p})~,
\end{equation}
where $\epsilon_{\mu\nu}$ is a constant polarization tensor, and 
$x$ are the $p+1$ worldvolume coordinates.
The graviton is polarized along the brane worldvolume coordinates, 
i.e. $\epsilon_{\mu\nu} = 0, \mu,\nu= p+1,\dots,9$. We use the 
transverse traceless
gauge, i.e. $\partial_{\mu} h^{\mu\nu} = 0$ and $h_{\mu}^{\mu} = 0$, and
consider the s-wave mode of the graviton, i.e. it does
not vary along the directions of the $(8-p)$-dimensional sphere.
We expand to quadratic order in the supergravity action for the graviton.
The type-II supergravity equation for the graviton modes $h(U)$ is 
related to a minimally coupled scalar field~\cite{aio}.
More precisely, for certain polarization and momentum vectors the 
graviton equation is the minimal scalar equation       
\be
f''(U) + \frac{8-p}{U}f'(U) - k^2 \lambda U^{p-7} f(U) = 0 \ .
\label{eq}
\ee
In the following we will suppress the index structure of the 
stress-energy tensor.

We consider the equation (\ref{eq}) in the range $0 \leq U \leq U_0$. 
This equation is solved by Bessel functions and we choose the 
regular solution that is normalized to one at $U=U_0$. 
Substituting the solution into the supergravity action, 
the generating functional is obtained as a pure boundary term.
We construct the ``flux factor'' ${\cal F}_T$  
\be
{\cal F}_T = \frac{N^2}{\lambda^2}\left[f(U)U^{8-p}
  \partial_U f(U)\right]_0^{U_0} \ ,
\label{flux}
\ee
which yields the momentum-space two-point function
and we get that
\be
 \langle T(k) T(-k) \rangle = {\cal F}_T \ .
\ee
By choosing the $U_0$ independent non-analytic term in ${\cal F}_T$, 
and Fourier transforming we obtain the two-point function in 
$x$-space.  When  $p<5$ we get
\begin{equation}
\label{ttcorr}
     \langle T(x) T(0) \rangle = \frac{N^2}{\lambda^{(2-\alpha)}}
         \frac{1}{|x|^{2\alpha+p+1}} = 
      \frac{N^\alpha}{g_{YM}^{2(2-\alpha)}}\frac{1}
        {|x|^{2\alpha+p+1}}\ ,~~~~~~~\alpha=\frac{7-p}{5-p} \ .
\end{equation}
For $p=5$ the two-point function reads
\begin{equation}
\label{ttcorr5}
     \langle T(x) T(0) \rangle = \frac{N^2}{\lambda^{19/4}}\frac{1}{|x|^{5/2}}
       e^{-|x|/{\sqrt\lambda}}\ ,
\end{equation}
where we have only written the leading large $|x|$ behaviour. 
We have summarised the results in
Table 1. 

Two cases which will be of particular importance for the brane-world scenarios are $p=4$ and
$p=5$. It is therefore of interest to give some more details on the computation in these
two cases.
\vskip 0.7 cm
{\bf D4-branes}\\
The minimal scalar equation (\ref{eq}) reads
\be
f''(U) + \frac{4}{U}f'(U) - k^2 \lambda U^{-3} f(U) = 0 \ .
\label{eq4}
\ee
Define the coordinate $z= 2\left(\frac{\lambda}{U}\right)^{\frac{1}{2}}$ which has dimensions
of length.
The regular normalizable solution of (\ref{eq4}) is given by
\be
f(z) = (k z)^3 K_3 (k z) \ ,
\ee
where $K_3(k z)$ is the modified Bessel function. The relevant non-analytic piece of
the ``flux factor'' (\ref{flux}) reads 
\be 
{\cal F}_T = N^2 \lambda k^6 \log(k) \ .
\label{F4}
\ee 
The Fourier transform of (\ref{F4}) yields the result
\be
 \langle T(x) T(0) \rangle = \frac{N^3 g_{YM}^2}{|x|^{11}} \ .
\label{TT}
\ee

At the crossover regime between the supergravity description and the perturbative Yang-Mills
description $g_{YM}^2N |x|^{-1} \simeq 1$ we have the matching to the expected $U(N)$ gauge
theory result with $N^2$ degrees of freedom 
\be
 \langle T(x) T(0) \rangle = \frac{N^2}{|x|^{10}} \ .
\label{TT1}
\ee 
The matching agrees 
up to a numerical factor that is independent of the rank of the gauge
group $N$ and the Yang-Mills coupling $g_{YM}$.

\vskip 0.7 cm
{\bf D5-branes}\\
The minimal scalar equation (\ref{eq}) reads 
\be
f''(U) + \frac{3}{U}f'(U) - k^2 \lambda U^{-2} f(U) = 0 \ .
\label{eq5}
\ee
Define $U \lambda^{\frac{1}{2}} = \exp(z \lambda^{-\frac{1}{2}})$, 
where $z$ has dimensions of length.
The solutions of (\ref{eq5}) are given by
\be
f(z) = \exp\left[-z \lambda^{-\frac{1}{2}}\left(1 \pm \sqrt{1+\lambda k^2}
\right)\right] \ .
\label{sol5}
\ee
After the normalization the 
relevant non-analytic  piece of
the ``flux factor'' (\ref{flux}) reads 
\be 
{\cal F}_T = \frac{N^2}{\lambda^3}\sqrt{1+ \lambda k^2} \ .
\label{F5}
\ee 
Fourier transforming (\ref{F5}) yields (\ref{ttcorr5}).

At the crossover regime between the supergravity description and the perturbative Yang-Mills
description $g_{YM}^2N |x|^{-2} \simeq 1$ we have the matching to the expected $U(N)$ gauge
theory result with $N^2$ degrees of freedom 
\be
 \langle T(x) T(0) \rangle = \frac{N^2}{|x|^{12}} \ .
\label{TT2}
\ee 
As before, the matching agrees up to a numerical factor that is independent 
of the rank of the gauge group $N$ and the Yang-Mills coupling $g_{YM}$.

At a length scale $|x| \ll \frac{g_{YM}}{\sqrt{N}}$ the effective string coupling is
large 
and the weakly coupled description is given by the S-dual NS5-brane background.
The minimally coupled scalar equation in this background is again (\ref{eq5}) and
the result
(\ref{ttcorr5}) is obtained again.

\vskip 0.7 cm
{\bf D$p$-branes}\\
For general $p$ the correlator $\langle T(x) T(0) \rangle$ as given in Table 1, 
matches at the crossover region  between the supergravity description and 
the perturbative Yang-Mills
description $g_{YM}^2N |x|^{p-3} \simeq 1$ with the expected Yang-Mills result
\be
\langle T(x) T(0) \rangle = \frac{N^2}{|x|^{2(p+1)}} \ .
\ee
We note that the result in Table 1 for the stress-energy two-point function
when $p=1$ was obtained in \cite{hash}.

\begin{table}
\begin{center}
\begin{tabular}{|c||c|}
\hline
\rule[-4mm]{0mm}{10mm}
$p$ & $\langle T(x) T(0) \rangle$ \\
\hline\hline
\rule[-5mm]{0mm}{11mm}
0& $\frac{N^{7/5}}{g_{YM}^{6/5}} \frac{1}{|x|^{19/5}}$ \\\hline
\rule[-4mm]{0mm}{10mm}
1& $\frac{N^{3/2}}{g_{YM}} \frac{1}{|x|^5}$ \\\hline
\rule[-5mm]{0mm}{11mm}
2& $\frac{N^{5/3}}{g_{YM}^{2/3}} \frac{1}{|x|^{19/3}}$ \\\hline
\rule[-4mm]{0mm}{10mm}
3& $\frac{N^{2}}{|x|^{8}}$ \\
\hline
\rule[-4mm]{0mm}{11mm}
4& $\frac{N^3 g_{YM}^2}{|x|^{11}}$ \\
\hline
\rule[-5mm]{0mm}{12mm}
5&  $\frac{N^2}{\lambda^{19/4}}\frac{1}{ |x|^{5/2}}e^{-|x|/\sqrt{\lambda}}$  \\
\hline
\end{tabular}
\end{center}
\caption{The supergravity results
for the $N$ D$p$-branes $\langle T(x) T(0)\rangle$ correlator.} 
\end{table}
\subsection{Current-current correlators}

Global symmetry currents of the brane worldvolume field theory $J_{\mu}$
couple to massless bulk gauge fields $A^{\mu}$ as
\be
S_{int} \sim \int d^{p+1}x J_{\mu}A^{\mu} \ .
\ee
We may distinguish two types of such gauge fields. 
Those that arise from the isometries
of the space on which we reduce, in our case the $(8-p)$-dimensional sphere
in (\ref{metric}). These couple to $SO(9-p)$ R-symmetry currents of the 
D$p$-branes gauge theory.
We can have additional gauge fields that couple to other global symmetry
currents as, for instance, in the five-dimensional fixed points with $E_n$ global symmetry
analysed in \cite{boz}.
We will consider the former type in the following.

In order to compute the current correlators we need to solve the graviton 
equation for the components of the graviton with one index along the brane 
worldvolume and one index along the $(8-p)$-dimensional sphere.
Alternatively we can first reduce the supergravity metric (\ref{metric}) on the
$(8-p)$-dimensional sphere and solve the $p+2$-dimensional gauge field 
equation.
We will pursue the second procedure since it will also shed light later on
brane-world scenarios based on various domain-wall solutions in the literature.

The dimensional reduction of the D$p$-brane supergravity solutions 
along the sphere have been performed in \cite{youm,clp} using 
another coordinate system. We will repeat the same procedure here, except
that we will use the coordinates of the supergravity solution (\ref{metric}).
It is useful to retain the $U$ coordinate because it has the 
interpretation of an energy scale of the brane worldvolume field theory.

The dimensional reduction is performed by using the ansatz
\be
   ds_E^2 = l_s^2\left( e^{a\varphi} ds_{p+2}^2 
   + e^{-\frac{p}{8-p} a\varphi}d\Omega_{8-p}^2\right) \, 
\label{reduced}
\ee
where $\varphi$ is a scalar field and $ds_E^2$ is the 
ten-dimensional D$p$-branes metric written in the Einstein frame, i.e.
it is related to the string frame metric (\ref{metric}) by
$ds_E^2 = e^{-\frac{\phi}{2}}ds^2$. The constant $a$ is given by
\be
a = \frac{1}{2}\left(\frac{8-p}{p}\right)^{\frac{1}{2}} \ .
\ee
By comparing the coefficient of the $d\Omega_{8-p}^2$ term of 
the two metrics $ds_E^2$ in (\ref{reduced})
and $ds^2$ in (\ref{metric}) we can read the form of the scalar $\varphi$ 
\be
e^{\varphi} = \left[\frac{1}{N} \lambda^{\frac{1}{4}(3-p)}
U^{-\frac{1}{4}(3-p)^2}\right]^{\left(\frac{8-p}{p}\right)^{\frac{1}{2}}} \ .
\end{equation}
Substituting this expression into (\ref{reduced}) we obtain 
the $(p+2)$-dimensional metric
\be 
ds_{p+2}^2 = N^{\frac{4}{p}}\lambda^{-\frac{3}{p}}U^{\frac{9}{p}-1}
\left(dx_{||}^2 + \lambda U^{p-7} dU^2 \right) \ .
\ee
The $(p+2)$-dimensional gauge field bulk action is then given by 
\be
S_{gauge} = \int d^{p+2}x \sqrt{g_{p+2}}\, 
   e^{-\frac{4}{\sqrt{p(8-p)}}\varphi}F^2
   \equiv \int d^{p+2}x \sqrt{g_{p+2}}\, \frac{1}{g^2_{bulk}}F^2 \ ,
\label{gauge}
\ee
where 
\be
\label{g2def}
g^2_{bulk} = N^{-\frac{4}{p}}\lambda^{\frac{3}{p}-1}U^{-\frac{(3-p)^2}{p}} \ .
\ee
This is of importance since while the gravitational and scalar fields in 
the reduced action are canonical in $(p+2)$ dimensions this is not the 
case for the gauge field as we see from (\ref{gauge}).
Note that for our purposes we omit the group theory factors 
and we write the action (\ref{gauge}) schematically.
In the following we will suppress the Lorentz index on the current

In order to solve the $(p+2)$-dimensional Maxwell equation, we
write the $(p+2)$-dimensional gauge field as
\be
A_{\mu}(x,U) = \varepsilon_{\mu}e^{ik \cdot x} f(U)   \ .
\ee
Imposing the gauge conditions $\partial^{\mu}A_{\mu} = 0, \mu=0,...,p+1$ and
$A_{U} = 0$ we obtain
\be
f''(U) + \frac{3}{U}f'(U) - k^2 \lambda U^{p-7} f(U) = 0 \ .
\label{eqa}
\ee

As before, we consider equation (\ref{eqa}) in the range $0 \leq U \leq U_0$. 
We solve this equation and choose a regular solution that is normalized
to one at $U=U_0$. 
The generating functional for the current correlators is obtained as a pure boundary term.
Again we construct the ``flux factor'' ${\cal F}_A$  
\be
{\cal F}_A = \frac{N^2}{\lambda}\left[f(U)U^{3}\partial_U f(U)\right]_0^{U_0} \ ,
\ee
and we have that
\be
 \langle J(k) J(-k) \rangle = {\cal F}_A \ .
\ee
By choosing the $U_0$ independent non-analytic term in ${\cal F}_A$,
and Fourier transforming we obtain for $p<5$
\begin{equation}
\label{jjcorr}
     \langle J(x) J(0) \rangle = \frac{N^2}{\lambda^{(2-\alpha)}}\frac{1}
        {|x|^{2\alpha+p-1}}= 
       \frac{N^\alpha}{g_{YM}^{2(2-\alpha)}}\frac{1}
        {|x|^{2\alpha+p-1}}\ ,~~~~~~~\alpha=\frac{7-p}{5-p} \ ,
\end{equation}
and for $p=5$
\begin{equation}
\label{jjcorr5}
     \langle J(x) J(0) \rangle = \frac{N^2}{\lambda^{15/4}}\frac{1}{|x|^{5/2}}
       e^{-x/\sqrt{\lambda}}\ .
\end{equation}
The results are summarized in Table 2.
Consider in detail the $p=4$ and $p=5$ cases.
\vskip 0.7 cm
{\bf D4-branes}\\
Equation (\ref{eqa}) reads 
\be
f''(U) + \frac{3}{U}f'(U) - k^2 \lambda U^{-3} f(U) = 0 \ .
\label{eq44}
\ee
As before, define $z= 2\left(\frac{\lambda}{U}\right)^{\frac{1}{2}}$ which has dimensions of length.
The regular normalized solution of (\ref{eq44}) is given by
\be
f(z) = (k z)^2 K_2 (k z) \ ,
\ee
where $K_2(k z)$ is the modified Bessel function. The relevant piece of
the ``flux factor'' (\ref{flux}) reads 
\be 
{\cal F}_A = N^2 \lambda k^4 \log(k) \ .
\label{F44}
\ee 
Fourier transforming (\ref{F44}) yields
\be
 \langle J(x) J(0) \rangle = \frac{N^3 g_{YM}^2}{|x|^{9}} \ .
\label{JJ}
\ee

At the crossover region between the supergravity description and the perturbative Yang-Mills
description $g_{YM}^2N |x|^{-1} \simeq 1$ we have the matching to the expected gauge
theory result
\be
 \langle J(x) J(0) \rangle = \frac{N^2}{|x|^{8}} \ .
\label{JJ1}
\ee 
The matching is up to a numerical factor that is independent of 
the rank of the gauge
group $N$ and the Yang-Mills coupling $g_{YM}$.
We have omitted the group theory factors and R-charges as well.

\vskip 0.7 cm
{\bf D5-branes}\\
Equation (\ref{eqa}) is identical to the minimal scalar equation (\ref{eq5}). 
We consider the same solutions (\ref{sol5}) and construct the ``flux factor'' 
which now reads
\be 
{\cal F}_A = \frac{N^2}{\lambda^2}\sqrt{1+\lambda k^2} \ .
\label{F55}
\ee 
Fourier transforming (\ref{F55}) yields (\ref{jjcorr5}).

At the crossover region between the supergravity description and the perturbative Yang-Mills
description $g_{YM}^2N |x|^{-2} \simeq 1$ we have the matching to the expected gauge
theory result 
\be
 \langle J(x) J(0) \rangle = \frac{N^2}{|x|^{10}} \ .
\label{JJ2}
\ee 
As before, 
the matching is up to a numerical factor that is independent of the rank of the gauge
group $N$ and the Yang-Mills coupling $g_{YM}$.

Similarly as for the stress-energy tensor the result (\ref{jjcorr5}) 
is also valid for large string coupling. This is because the S-dual
supergravity background of $N$ NS-5 branes leads to the same equation 
of motion as (\ref{eq5}).

\vskip 0.7 cm
{\bf D$p$-branes}\\
Note that the correlator $\langle J(x) J(0) \rangle$ for general $p$, as given in Table 2, 
matches at the crossover region  between the supergravity description and 
the perturbative Yang-Mills
description $g_{YM}^2N |x|^{p-3} \simeq 1$ with the expected Yang-Mills result
\be
\langle J(x) J(0) \rangle = \frac{N^2}{|x|^{2p}} \ .
\ee
Note also that, up to numerical constants that are independent of $g_{YM}$ and $N$, the two-point functions
of the stress-tensor $T$ and the current $J$ are related for $p < 5$ by
\be
\langle J(x) J(0) \rangle = |x|^{2}\langle T(x) T(0) \rangle \ ,
\ee
and for $p=5$ by
\be
\langle J(x) J(0) \rangle = \lambda\langle T(x) T(0) \rangle \ .
\ee

\begin{table}
\begin{center}
\begin{tabular}{|c||c|}
\hline
\rule[-4mm]{0mm}{9mm}
$p$ & $\langle J(x)J(0) \rangle$ \\
\hline\hline
\rule[-5mm]{0mm}{11mm}
0& 
$\frac{N^{7/5}}{g_{YM}^{6/5}} \frac{1}{|x|^{9/5}}$ \\\hline
\rule[-4mm]{0mm}{10mm}
1& 
$\frac{N^{3/2}}{g_{YM}} \frac{1}{|x|^3}$ \\\hline
\rule[-5mm]{0mm}{11mm}
2& 
$\frac{N^{5/3}}{g_{YM}^{2/3}} \frac{1}{|x|^{13/3}}$ \\\hline
\rule[-4mm]{0mm}{10mm}
3& $\frac{N^{2}}{|x|^{6}}$ \\
\hline
\rule[-4mm]{0mm}{11mm}
4& $\frac{N^3 g_{YM}^2}{|x|^9}$ \\
\hline
\rule[-5mm]{0mm}{12mm}
5& $\frac{N^2}{\lambda^{15/4}}\frac{1}{|x|^{5/2}} e^{-|x|/\sqrt{\lambda}} $ \\
\hline
\end{tabular}
\end{center}
\caption{The supergravity results
for the $N$ D$p$-branes $\langle J(x) J(0)\rangle$ correlator.} 
\end{table}

\section{Brane-worlds}

We use the notation brane-world for a brane (not necessarily a  D$p$-brane) located at $U=U_{*}$.
When $p\neq 3$ and the brane is a  D$p$-brane, 
the theory on the brane-world is described by a $(p+1)$-dimensional
non-conformal super Yang-Mills theory coupled to gravity
with a cutoff $\Lambda = U_{*}$.
In general, the brane-world can have extra degrees of freedom.
 
In this section we will analyse these brane-world scenarios with and without additional
compactification of brane worldvolume coordinates.
These brane-world scenarios are sometimes called dilatonic domain walls.
To make the connection we note that
in $p+2$ dimensions the RR $(p+2)$-form field $F_{p+2}=dA_{p+1}$ can be dualized and replaced
by a cosmological constant. The $p+2$-dimensional supergravity action reduced
from ten-dimensions has domain wall solutions. Some of these supersymmetric 
solutions are the ten-dimensional D$p$-branes solutions reduced on 
the $(8-p)$-dimensional sphere \cite{Boonstra:1999mp,Behrndt:1999mk,clp}.

A further 
reduction to be discussed is the compactification of some of the D$p$-branes
worldvolume coordinates. In particular the compactification of one coordinate
of the D4-brane worldvolume on the circle, $S^1$ 
or two of the D5-brane worldvolume on a torus, $S^1\times S^1$ 
leads to a four-dimensional brane-world scenarios.
We will discuss in more detail the relation to the brane-world scenarios based on
the dilatonic domain wall solutions in the discussion section.

\subsection{Newton's law}

The $(p+1)$-dimensional gravitational coupling $\kappa_{p+1}$ is 
defined in terms of the ten-dimensional gravitational coupling 
$\kappa_{10}$ via
\be
\frac{1}{\kappa_{10}^2} \int d^{10}x \sqrt{g_{10}}e^{-2 \phi}{\cal R}_{10} = 
\frac{1}{\kappa_{p+1}^2} \int d^{p+1}x \sqrt{g_{p+1}}{\cal R}_{p+1} \ .
\label{def}
\ee
The LHS is the ten-dimensional gravitational action in the string frame metric (\ref{metric})
while the RHS is the 
$(p+1)$-dimensional gravitational action in the Einstein  frame.
Using (\ref{metric}) and (\ref{def}) we obtain
\be
G_{p+1} \equiv \kappa_{p+1}^2 = \frac{g_{YM}^2}{N U_{*}^2} \ ,
\label{G}
\ee
where $U_{*}$ is the UV cutoff.
Note that when  $U_{*} \rightarrow \infty$, the gravitational coupling (\ref{G}) vanishes and gravity
decouples from the brane, as expected.

Consider the correction to Newton's law in the brane-world scenarios.
\vskip 0.7 cm
{\bf D4-branes}\\
Consider a brane-world scenario described by a five-dimensional SYM theory coupled to gravity
with a cutoff at $U=U_{*}$.
The correction to the Newton's law 
can be computed from the graviton propagator $G_{graviton}(k)$~\cite{gubser},
as in Figure 1.
The first part is the free graviton propagation while the second part is the leading non-CFT
correction
\bea
G_{graviton}(k) &\sim& \frac{1}{k^2} + \frac{1}{k^2}\kappa_{5}
\langle T(k) T(-k) \rangle \kappa_{5}\frac{1}{k^2} \nonumber\\
 &=& \frac{1}{k^2}\left(1+ N^3 g_{YM}^2G_{5}k^4\log(k)\right) \ ,
\eea
where we have used (\ref{F4}), and $G_5$ is given by (\ref{G}).
Fourier transforming in five dimensions yield
\be
\label{G(x)}
  G_{graviton}(x) \sim \frac{1}{x^3}\left(1  + 
    N^3g_{YM}^2 \frac{G_{5}}{x^4} \right) \ .
\ee
From (\ref{G(x)}) we can read 
the modification to the five-dimensional Newton's force law $\frac{1}{r^3}$,
\be
\label{forceD4}
F_{Newton} = \frac{G_5m_1m_2}{r^3} \left(1  + 
    \frac{N^2\lambda G_5}{r^4}  \right) \ .
\ee

\vskip 0.7 cm
{\bf D5-branes}\\
Consider  a six-dimensional brane-world scenario.
The correction to the Newton's law 
can be computed as before from the graviton propagator $G_{graviton}(k)$ as in
Figure 1. Again,
the first part is the free graviton propagation while the second part is the leading non-CFT
correction. Using (\ref{F5}) we get 
\be
G_{graviton}(k) \sim 
   \frac{1}{k^2}\left(1+ \frac{N^2 G_6}{\lambda^3}
   \frac{\sqrt{1+\lambda k^2}}{k^2}
\right) \ ,
\ee
where $G_6$ is given by (\ref{G}).
Fourier transforming in six dimensions yields at large $x$
\be
\label{gravx}
  G_{graviton}(x) \sim \frac{1}{x^4}\left(1  + 
    c\frac{N^2G_6}{\lambda^2} \left(\frac{x}{\sqrt{\lambda}}\right)^{\frac{3}{2}}e^{-x/\sqrt{\lambda}}
 \right) \ .
\ee
where by $c$ we denote a numerical constant which we neglected.
The modification to the six-dimensional Newton's force law $\frac{1}{r^4}$
reads
\be
\label{forcen5}
F_{Newton} = \frac{G_6m_1m_2}{r^4} \left(1  + c 
    \frac{N^2G_6}{\lambda^2} \left(\frac{r}{\sqrt{\lambda}}\right)^{\frac{3}{2}}e^{-r/\sqrt{\lambda}}
 \right) \ .
\ee

\begin{figure}[htb]
\begin{center}
\epsfxsize=4in\leavevmode\epsfbox{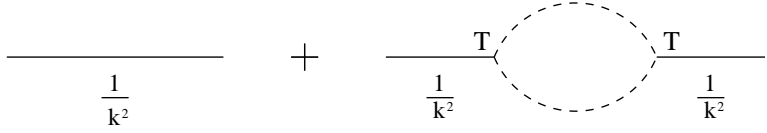}
\end{center}
\caption{\it The graviton propagator $G_{graviton}(k)$ 
where the first part is the free graviton propagation while the second 
part is the leading non-CFT correction.}
\label{Tfig}
\end{figure}

\vskip 1cm
{\bf Dimensional reduction}\\
In order to arrive at four-dimensional brane-world 
scenarios one can further compactify the worldvolume coordinates.
In the D4-brane case we compactify one coordinate on a circle while
in the D5-brane case we compactify two coordinates on a torus.
Consider first the D4-branes scenario where we compactify on a circle 
of radius $R$. At distances $|x| = U^{-1}$ in the regime  
\be
R \ll |x| \ll g_{YM}^2N \ ,
\label{regime1}
\ee
we can neglect the 
momenta along the circle direction and we are still in the supergravity regime
defined by (\ref{regime}). Thus, Fourier transforming in four dimensions,
the gravitational force is then
\be
\label{forceD44}
F_{Newton} = \frac{G_4m_1m_2}{r^2} \left(1  + 
    \frac{N^2\lambda G_4R}{r^4}  \right) \ ,
\ee
where $G_4 R=G_5$.

Consider next the D5-brane scenario where we compactify on two circles 
of radius $R$. At distances in the regime  
\be
R \ll |x| \ll (g_{YM}^2 N)^{\frac{1}{2}} \ ,
\label{regime2}
\ee
we can neglect the 
momenta along the circle direction and we are still in the supergravity regime
defined by (\ref{regime}).
Fourier transforming in four dimensions,
the gravitational force reads
\be
\label{forced54}
F_{Newton} = \frac{G_4m_1m_2}{r^2} \left(1  + c 
    \frac{N^2G_4R^2}{\lambda^2} \left(\frac{r}{\sqrt{\lambda}}\right)^{-\frac{1}{2}}e^{-r/\sqrt{\lambda}}
 \right) \ ,
\ee
where $G_4 R^2=G_6$.

\subsection{Coulomb's law}

The effective $(p+1)$-dimensional gauge coupling, $g_{brane}$, on the $p$-brane
can be obtained via
\be
 \int d^{p+2}x \sqrt{g_{p+2}} \frac{1}{g^2_{bulk}} F_{(p+2)}^2 = 
 \int d^{p+1}x \frac{1}{g_{brane}^2} F_{(p+1)}^2 \ ,
\label{gdef}
\ee
where $F_{(p+2)}$ and $F_{(p+1)}$ denote the field strength of the gauge field in $p+2$ and
$p+1$ dimensions, respectively. 
The LHS is the $(p+2)$-dimensional gauge field action after the dimensional
reduction (\ref{reduced})
while the RHS is the effective $(p+1)$-dimensional gauge field action.
Using (\ref{g2def}) we obtain for $p>3$
\be
  g_{brane}^2 = \frac{1}{N^2 U_{*}^{p-3}} \ ,
\label{gg}
\ee
where $U_{*}$ is the UV cutoff.

Consider the correction to Coulomb's law in these brane-world scenarios.
\vskip 0.7 cm
{\bf D4-branes}\\
Consider a five-dimensional brane-world scenario.
The correction to the Coulomb's law 
can be computed from the propagator for a vector field $G_{vector}(k)$ as in
Figure 2.
The first part is the free vector propagation while the second part is the leading non-CFT
correction
\bea
G_{vector}(k) &\sim& 
  \frac{1}{k^2} + \frac{1}{k^2}g_{5}\langle J(k)J(-k) \rangle
  g_{5}\frac{1}{k^2} \nonumber\\
&=& \frac{1}{k^2}\left(1+ N^3 g_{YM}^2 g_5^2 k^2 \log(k)\right) \ ,
\eea
where we have used (\ref{F44}). 
The gauge coupling $g_5$ is given by (\ref{gg}) for $p=4$.

Fourier transforming in five dimensions yields
\be
G_{vector}(x) \sim \frac{1}{x^3}\left(1+ N^3g_{YM}^2 \frac{g_5^2}{x^2}
\right) \ .
\ee
The modification to the five-dimensional Coulomb's force law $\frac{1}{r^3}$
reads
\be
\label{correction}
F_{Coulomb} = \frac{g_5^2 q_1 q_2}{r^3} \left(1 + \frac{N^2 \lambda g_5^2}{r^2}\right) \ .
\ee

\vskip 0.7 cm
{\bf D5-branes}\\
Consider a six-dimensional brane-world scenario.
The correction to the Coulomb's law 
can be computed as before from the vector propagator $G_{vector}(k)$ as in
Figure 2. Again, the first part is the free graviton propagation while 
the second part is the leading non-CFT
correction. Using (\ref{F55}) we obtain 
\be
G_{vector}(k) \sim 
\frac{1}{k^2}\left(1+ \frac{N^2 g_6^2}{\lambda^2} 
    \frac{\sqrt{1+\lambda k^2}}{k^2}
\right) \ ,
\ee
where  $g_6$ is given by (\ref{gg}) for $p=5$.
The modification to the six-dimensional Coulomb's force law $\frac{1}{r^4}$
reads
\be
\label{Cforce5}
F_{Coulomb} = \frac{g_6^2q_1q_2}{r^4} \left(1  + c 
    \frac{N^2g_6^2}{\lambda} \left(\frac{r}{\sqrt{\lambda}}\right)^{\frac{3}{2}}e^{-r/\sqrt{\lambda}}
 \right) \ .
\ee

\begin{figure}[htb]
\begin{center}
\epsfxsize=4in\leavevmode\epsfbox{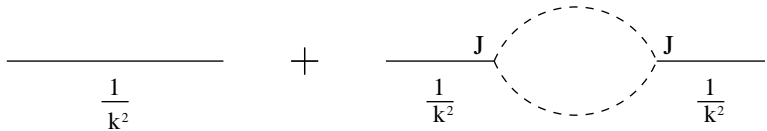}
\end{center}
\caption{\it The vector field propagator $G_{vector}(k)$ 
where the first part is the free vector propagation while the second 
part is the leading non-CFT correction.}
\label{Jfig}
\end{figure}

\vskip 1cm
{\bf Dimensional reduction}\\
Again, in order to arrive at four-dimensional brane-world 
scenarios one can further compactify the 
worldvolume coordinates.
In the D4-brane case we compactify one coordinate on a circle while
in the D5-brane case we compactify two coordinates on a torus.
Consider first the D4-branes scenario where we compactify on a circle 
of radius $R$. At distances satisfying (\ref{regime1})
the Coulomb force reads
\be
\label{Cforce44}
F_{Coulomb} = \frac{g_4^2 q_1 q_2}{r^2} \left(1 + \frac{N^2\lambda g_4^2R}{r^2}\right) \ .
\ee
where $g_4^2 R = g_{5}^2$.

Consider next the D5-branes scenario where we compactify on two
circles of radius $R$.
At distances satisfying (\ref{regime2}) we get

\be
\label{Cforce54}
F_{Coulomb} = \frac{g_4^2q_1q_2}{r^2} \left(1  + c 
    \frac{N^2g_4^2R^2}{\lambda} \left(\frac{r}{\sqrt{\lambda}}\right)^{-\frac{1}{2}}e^{-r/\sqrt{\lambda}}
 \right) \ .
\ee
where $g_4^2 R^2 = g_{6}^2$.

\subsection{Graviton and gauge field localisation}

In the previous calculation of the Newton and Coulomb laws, a massless
mode with behaviour $1/k^2$ was assumed to be localised on the
brane worldvolume. However, one can explicitly check the localisation 
properties of the bulk graviton and gauge fields on the $p$-brane 
located at $U=U_{*}$ by solving the equation of motion in the 
background geometry (\ref{metric}), and computing the Green's 
function ${\cal G}_k$. 

By invoking the bulk-boundary
correspondence, the bulk Green's function can be used to obtain
an alternative derivation of the two-point functions 
$\langle {\cal O} {\cal O}\rangle$ in the dual theory, 
considered in Section 2.
Since the Green's function represents the sum over all connected 
diagrams we obtain the relation 
\begin{equation}
\label{tpf}
     \langle {\cal O}(k) {\cal O}(-k)\rangle = \frac{k^2}{\kappa} - 
        {\cal G}_k^{-1}~,
\end{equation}
where for the stress-energy tensor $T_{\mu\nu}$ we use the graviton 
propagator $(\kappa=-2G_{p+1})$,
while for the vector current $J_\mu$ we use the gauge field 
propagator $(\kappa=(3-p)g_{p+1}^2)$.
For $p<6$ there always exists a term in (\ref{tpf})
which does not depend on the cutoff $U_*$. The results for the
stress-energy tensor and current correlator, obtained in this way, are
all in agreement with those found in section 2.

Let us now consider the graviton and gauge field Green's function 
to explicitly check the localisation properties on the $p$-brane.

\subsubsection{Graviton}

Consider a bulk graviton in the effective $p+2$-dimensional brane-world.
The minimally coupled scalar equation (\ref{eq}) determines the mass 
and localisation properties of the $p+1$-dimensional Kaluza-Klein modes, 
where $k^2=-m^2$. The differential operator (\ref{eq}) is self-adjoint and 
the solutions $f_m$ form a complete set which satisfy the orthonormal relation
\begin{equation}
     \int_0^{U_*} dU U f_m f_{m'} = \delta_{mm'}~.
\end{equation}
When $m=0$ we find that the equation of motion (\ref{eq}) 
always has a solution $f_0(U) = \rm constant$, and there is always 
a normalisable mode. This is consistent with (\ref{G}).

To check whether gravity is actually localized on the $p$-brane we must 
compute the Green's function for two sources on the brane. In order to 
do this the first question that needs to be addressed are the 
boundary conditions. This is particularly important because except 
for the case $p=3$, there is singular behaviour at $U=0$. This 
singularity corresponds to the massless
open string degrees of freedom that are missing in the description. 
We expect the singularity to be resolved once the required degrees are added,
and we will continue to impose Hartle-Hawking boundary conditions at $U=0$, 
as in the $p=3$ case. Thus following the procedure outlined in the Appendix the
expression for the scalar Green's function when $U=U'=U_{*}$ is given by
\begin{equation}
\label{gravgf}
    {\cal G}_k(U_{*}, U_{*}) =
        -i\frac{g_{YM}^4}{\sqrt{\lambda} k} U_{*}^{\frac{1}{2}(p-9)} 
       \frac{H_\alpha^{(1)} \left( i(\alpha-1)\sqrt{\lambda} k 
       U_{*}^{\frac{1}{2}(p-5)}\right)}{H_{\alpha-1}^{(1)}
        \left( i (\alpha-1)\sqrt{\lambda} k U_{*}^{\frac{1}{2}(p-5)}\right)}~,
\end{equation}
where $H^{(1)}$ is the Hankel function of the first kind. When 
$p<5$ the Green's function can be written in the form 
${\cal G}_k(U_{*}, U_{*}) = {\cal G}_k^{(0)} + {\cal G}_k^{(KK)}$,
where the zero mode contribution ${\cal G}_k^{(0)}$ and the Kaluza-Klein 
contribution ${\cal G}_k^{(KK)}$ are given by
\begin{eqnarray}
       {\cal G}_k^{(0)} &=& -2\frac{G_{p+1}}{k^2}~,\\
       {\cal G}_k^{(KK)} &=&i\frac{g_{YM}^4}{\sqrt{\lambda} k} U_{*}^{\frac{1}{2}(p-9)} 
        \frac{H_{\alpha-2}^{(1)}
        \left(i(\alpha-1)\sqrt{\lambda} k U_{*}^{\frac{1}{2}(p-5)}\right)}
       {H_{\alpha-1}^{(1)}
       \left(i(\alpha-1)\sqrt{\lambda} k U_{*}^{\frac{1}{2}(p-5)}\right)}~.
\end{eqnarray}
We see from the form 
of ${\cal G}_k^{(0)}$ that one obtains Newton's law 
on the $p$-brane located at $U_*$. This is consistent with the fact 
that the graviton zero mode is also localised there.
The corresponding corrections to the $(p+1)$-dimensional Newton's law 
are obtained from ${\cal G}_k^{(KK)}$ and one obtains the
result found earlier for the D4-brane.

When $p=5$ the Green's function has the form
\begin{equation}
\label{d5gf}
      {\cal G}_k(U_{*},U_{*}) = -\frac{G_6}{k^2} 
      \left( 1+\sqrt{1+\lambda k^2}\right)~,
\end{equation}
and again there is a localised massless mode $(1/k^2)$ at $U=U_*$.
The corresponding corrections to Newton's law arising from the
Kaluza-Klein continuum lead to exponentially suppressed corrections, and
agrees with the result found earlier.

When $p\geq 6$ one finds that there is an equal and opposite 
contribution arising from ${\cal G}_k^{(KK)}$ that precisely cancels 
the ${\cal G}_k^{(0)}$ contribution.
The remaining leading contribution from ${\cal G}_k^{(KK)}$
does not lead to the usual Newton's law. Thus for $p\geq 6$ we do not 
obtain Newton's law and there is no localised mode in the brane-world
with $1/k^2$ behaviour.

\subsubsection{Gauge fields}

The equation of motion for the $(p+1)$-dimensional Kaluza-Klein
gauge fields is governed by (\ref{eqa}) where $k^2=-m^2$.
The differential operator in (\ref{eqa}) is self-adjoint and the
solutions $f_m$ form a complete set which satisfy 
the orthonormal relation
\begin{equation}
   \int_0^{U_*} dU U^{p-4} f_m f_{m'} = \delta_{mm'}~.
\end {equation}
Notice that when $m=0$ there is always a constant solution to Eq.~(\ref{eqa}).
Thus only for $p\geq4$ will the zero mode be normalisable. This is
consistent with (\ref{gg}).


Analogous to the graviton case we can obtain the Green's function for 
two charged (under the gauge field) sources localised on the $p$-brane.
Following the procedure outlined in the Appendix one obtains the
gauge field Green's function
\begin{eqnarray}
\label{gaugegf}
    {\cal G}_k(U_{*}, U_{*}) &=&
        -i\frac{\sqrt{\lambda}}{N^2 k} U_*^{\frac{1}{2}(1-p)} 
    \frac{H_{\alpha-1}^{(1)}
        \left(i(\alpha-1)\sqrt{\lambda} k U_*^{\frac{1}{2}(p-5)}\right)}{H_{\alpha-2}^{(1)}
        \left(i(\alpha-1)\sqrt{\lambda} k U_*^{\frac{1}{2}(p-5)}\right)} \\
    &\equiv& {\cal G}_k^{(0)} (U_*) + {\cal G}_k^{(KK)}(U_{*})\nonumber~,
\end{eqnarray}
where for $p\neq 5$ we have
\begin{eqnarray}
\label{gfgf0}
       {\cal G}_k^{(0)} &=& (3-p) \frac{g_{p+1}^2}{k^2}~,\\
\label{gfgfKK}
       {\cal G}_k^{(KK)} &=& i\frac{\sqrt{\lambda}}{N^2 k} 
     U_*^{\frac{1}{2}(1-p)} \frac{H_{\alpha-3}^{(1)}
      \left(i(\alpha-1)\sqrt{\lambda} k U_*^{\frac{1}{2}(p-5)}\right)}
    {H_{\alpha-2}^{(1)}
      \left(i(\alpha-1)\sqrt{\lambda} k U_*^{\frac{1}{2}(p-5)}\right)}~,
\end{eqnarray}
and $g_{p+1}=1/(N U_*^{(p-3)/2})$ is the effective gauge coupling for $p>3$
on the $p$-brane.
When $p<3$, the leading contribution ${\cal G}_k^{(0)}$ is positive
and cancels against an equal and opposite contribution
from ${\cal G}_k^{(KK)}$. Thus, there is no localised mode with $1/k^2$
behaviour.

For $p=3$, the zero mode contribution ${\cal G}_k^{(0)}$ vanishes and the 
leading Kaluza-Klein contribution is 
\begin{equation}
    {\cal G}_k^{(KK)} = \frac{1}{N^2 \log(k)} 
   \frac{1}{k^2} \equiv \frac{e_{eff}^2(k)}{k^2}~.
\end{equation}
As noticed in \cite{pomarol,kss,apr}, even though the bulk zero mode
is nonnormalizable, one can still think of a 4d gauge field with an effective
gauge coupling that runs logarithmically to zero in the infrared.

When $p=4$ the zero mode contribution ${\cal G}_k^{(0)}$ 
is non-vanishing and there is a localised gauge field on the brane. 
The corrections to Coulomb's law arise from the Kaluza-Klein continuum
and agrees with the result obtained earlier for the D4-brane.

The D5-brane case is similar to that encountered for the minimally 
coupled scalar. The Green's function is now
\begin{equation}
\label{gfgf5}
      {\cal G}_k(U_*,U_*) = -\frac{g_6^2}{k^2} \left( 
      1+\sqrt{1+\lambda k^2}\right)~,
\end{equation}
and one obtains the six-dimensional Coulomb's law.
These results are identical to the graviton case except for the 
couplings and again we see the mass gap in the Kaluza-Klein spectrum 
of the gauge field.

When $p>5$ the contribution ${\cal G}_k^{(0)}$ is again cancelled by an 
equal and opposite contribution arising from ${\cal G}_k^{(KK)}$. Thus 
there is no localised gauge field on the $p$-brane with $1/k^2$ behaviour.

\section{Discussion}

In the following we will discuss in more detail the relation between our work
and the brane-world scenarios based on dilatonic domain wall solutions.
Consider domain wall solutions in $D$-dimensions
which are solutions of the field equations derived from the action
\begin{equation}
    S= \int d^{D}x\, \sqrt{g}\left[{\cal R} 
      -\frac{1}{2}(\partial\varphi)^2 - 2 \Lambda e^{b\varphi}\right]
      + \int d^{D-1}x\, {\cal L} \ .
\end{equation}
The field $\varphi$ is a $D$-dimensional scalar field, 
$\Lambda$ is a bulk cosmological constant and 
$b$ is a constant parameterised in terms of a quantity $\Delta$
\begin{equation}
    b^2 = \Delta + \frac{2(D-1)}{(D-2)} \ .
\end{equation}
The lagrangian ${\cal L}$ is a delta-function source added to the bulk action. It provides a cutoff for the
the boundary of the bulk solution.

When $b=0$ one obtains AdS solutions where the dilaton is constant
and the dual description is some CFT.
More precisely, if we use only a slice of AdS, the description of the brane-world scenario is
in terms of a CFT coupled to gravity with a cutoff.
For example, when $D=5$  then $\Delta_{\rm AdS_5} = -8/3$, and 
one obtains the 
solution~\cite{rsII}. This can originate, for instance, from a ten-dimensional
$AdS_5\times S^5$ solution
of type-IIB supergravity, with an 
${\cal N}=4$ SCFT dual description.
In the truncated brane-world version, 
the dual conformal field theory is modified 
by introducing an ultraviolet cutoff, and gauging the Poincare 
symmetry \cite{apr,rz}.

Similarly, when $D=6$ one obtains a  six-dimensional 
AdS solution ~\cite{gs}, where $\Delta_{\rm AdS_6} = -5/2$. 
This can arise as a reduction
of a ten-dimensional warped  AdS$_6$ solution
corresponding to the D4-D8 brane system~\cite{boz}. The dual field theory
is a five-dimensional CFT.
A five-dimensional domain wall solution can be obtained 
by compactifying one of the brane dimensions on a circle $S^1$, and   
analysed in terms 
of the corresponding dual theory in \cite{poppitz}. Other 
similar solutions were obtained in \cite{grs}.

Domain wall solutions can also be constructed where $b\neq 0$, and
the dilaton is no longer constant. One example is in $D=5$ and
$\Delta = -12/5$ with a varying dilaton \cite{Boonstra:1999mp,Behrndt:1999mk,youm,clp}, which
can be obtained from the near-horizon metric of
D4-branes 
compactified on $S^1\times S^4$. 
Following the results in section 3.1, the Newtonian law
obtained from the dimensional reduction of the
D4-brane solution is consistent with the corrections 
obtained in \cite{clp}. Our derivation indicates that
the dual theory of this domain wall solution is 
the five-dimensional SYM theory compactified on $S^1$.
In comparison to the Randall-Sundrum scenario where 
the correction to the Newton potential is proportional to $1/r^3$, here the corrections 
are $1/r^5$. 
We also considered  the gauge fields in this 
domain wall background. Gauge fields that arise from the 
isometries of $S^4$ have a 
zero mode localised on the domain wall.
The corrections to the Coulomb law arising from the Kaluza-Klein 
continuum are given by (\ref{correction}), and are related
to the current correlators of the dual
field theory.

Another $D=5$ solution, 
with $\Delta=-2$, arises
from the near-horizon metric of D5-branes compactified
on $T^2\times S^3$.
This case, analysed in section 3.1,
gives rise to a Newtonian force with 
exponentially suppressed corrections (see also \cite{clp}). 
Our analysis indicates 
that the dual field theory is a six-dimensional SYM compactified on $T^2$. 
The exponentially suppressed corrections have two equivalent
interpretations. In the supergravity description, they arise because
the Kaluza-Klein continuum is separated from the zero mode by a 
mass gap. In the six-dimensional  SYM theory they correspond to a mass gap of the theory developed
in
the strongly coupled regime.
As before, 
we can study the gauge fields in the
domain wall background. Those that arise from the isometries of $S^3$
have a zero mode 
localised on the domain wall. The Coulomb law corrections 
are exponentially suppressed, and they are 
related to the current
correlators of the six-dimensional SYM theory.

\section*{Acknowledgements}
The work of TG is supported by the FNRS, contract no. 21-55560.98.


\section*{Appendix: Green's function in the extremal D$p$-brane geometry}

We will follow the derivation of the Green's function presented 
in Ref.~\cite{gkr}, except that we will consider the extremal
D$p$-brane geometry. Let us first consider the case of the minimally coupled
scalar $\phi$ in ten dimensions. Introducing a source 
function $\cal J$, one obtains 
\begin{equation}
\label{geneqn}
     \partial_M( \sqrt{-g} g^{MN} \partial_N \phi(X)) = \sqrt{-g} {\cal J}(X)~. 
\end{equation}
where $X=(x,U,\theta_i)$, with $x$ denoting the $p+1$ worldvolume coordinates. 
The corresponding Green's function for (\ref{geneqn}) can then be defined as
\begin{equation}
    \phi(X) = \int d^{10} X^\prime \sqrt{-g} \,
       G(X; X^\prime) {\cal J}(X^\prime)~.
\end{equation}
If we now consider the Fourier transform of the Green's function
\begin{equation}
     G(X;X^\prime) = \sum_{\{n_i\}} e^{i \{n_i\}\theta_i} \int 
     \frac{d^{p+1} k}{(2\pi)^{p+1}} 
      e^{ik \cdot (x-x^\prime)} {\cal G}_k(U,U^\prime)~,
\end{equation}
where the discrete Fourier transform is over the hyperspherical coordinates of $S^{8-p}$,
then the Fourier component ${\cal G}_k(U,U^\prime)$ must satisfy the equation
\begin{equation}
\label{Gpeqn}
    \left[ \partial_u( U^{8-p} \partial_u) -\lambda k^2 U + m_{\theta_i}^2 U^{6-p} \right] 
      {\cal G}_k(U,U^\prime) = g_{YM}^4 \delta(U-U^\prime)~.
\end{equation}
The standard procedure for solving Eq.~(\ref{Gpeqn}) is to use 
the solution to the homogeneous equation in the regions $U<U^\prime$ 
and $U > U^\prime$, and then impose matching conditions at $U=U^\prime$.
If we now restrict to the case where $m_{\theta_i}^2=0$ and write
\begin{equation}
        {\cal G}_k (U,U^\prime) = \theta(U-U^\prime) {\cal G}_> +
           \theta(U^\prime - U) {\cal G}_<~,
\end{equation}
the solution to the homogeneous equation for $U >U^\prime$ and $p\neq 5$ is given by
\begin{eqnarray}
        {\cal G}_>(U,U^\prime) &=& i A_>(U^\prime) U^{\frac{1}{2}(p-7)}
       \left[ J_{\alpha -1}(q_*) H_\alpha^{(1)}(q)
         - H_{\alpha-1}^{(1)}(q_*) J_\alpha(q) \right]~,
\end{eqnarray}
with $\alpha=(7-p)/(5-p)$ and 
where we have imposed the Neumann condition $\partial_u {\cal G}_k(U,U^\prime)
\big |_{U=U_*} = 0$. We have also defined
$q= i(\alpha-1){\sqrt\lambda}k U^{(p-5)/2}$, and 
$H_\alpha^{(1)}$ is the Hankel function of the first kind of order $\alpha$. 
As noted earlier there is a singularity at $U=0$, but since the solution can
be made finite there we will continue to impose the Hartle-Hawking boundary 
condition. Thus for $U<U^\prime$ we obtain
\begin{equation}
      {\cal G}_<(U,U^\prime) = i A_<(U^\prime) U^{\frac{1}{2}(p-7)}
         H_\alpha^{(1)}(q)~.
\end{equation}
The unknown functions $A_<(U^\prime)$
and $A_>(U^\prime)$ are determined by imposing matching conditions
at $U=U^\prime$. Continuity of ${\cal G}_k$ at $U=U^\prime$ leads to
the condition
\begin{eqnarray}
   {\cal G}_> \big |_{U=U^\prime} = {\cal G}_< \big |_{U = U^\prime}~, 
\end{eqnarray}
while the discontinuity in $\partial_u {\cal G}_k$ gives the condition
\begin{eqnarray}
      \left(\partial_u {\cal G}_>-  \partial_u {\cal G}_< 
      \right) \bigg |_{U=U^\prime}= g_{YM}^4 U^{\prime p-6}~.
\end{eqnarray}
This leads to the solutions
\begin{eqnarray}
     A_<(U^\prime) &=&  \frac{\pi g_{YM}^4}{5-p} 
        U^{\prime \frac{1}{2}(p-7)}
      \frac{H_\alpha^{(1)}(q')}
          {H_{\alpha-1}^{(1)}(q_*)}~,\\
     A_>(U^\prime)& =&  i \frac{\pi g_{YM}^4}{5-p} 
        U^{\prime\frac{1}{2}(p-7)}
        \left[\frac{J_{\alpha-1}(q_*)}
               {H_{\alpha-1}^{(1)}(q_*)}
             H_\alpha^{(1)}(q')-J_\alpha(q')\right].
\end{eqnarray}
Finally, substituting these functions into the equations 
for ${\cal G}_>$ and ${\cal G}_<$ gives the expression for 
the Green's function in the extremal D$p$-brane geometry
\begin{eqnarray}
\label{greenfn}
     {\cal G}_k(U,U^\prime) &=&i \frac{\pi g_{YM}^4 }{5-p} 
        (U U^\prime)^{\frac{1}{2}(p-7)} 
       \frac{H_\alpha^{(1)}(q_<)}{H_{\alpha-1}^{(1)}(q_*)} \nonumber \\
       &&\qquad\qquad \times\left[J_{\alpha-1} (q_*) H_\alpha^{(1)}(q_>)
     - H_{\alpha-1}^{(1)}(q_*)J_\alpha(q_>) \right]~,
\end{eqnarray}
where in $q_> (q_<)$ we have defined $U_> (U_<)$ to be the greater (lesser) of $U$ and 
$U^\prime$. The Green's function (\ref{greenfn}) is the general expression 
when $p\neq5$
for a minimally coupled scalar in the extremal D$p$-brane geometry. Notice also
that from the pole condition $H_{\alpha-1}^{(1)} (2i {\sqrt{\lambda}} k/(5-p) 
U_*^{\frac{1}{2}(p-5)})= 0$, 
there is a branch cut singularity at $k=0$. This represents the Kaluza-Klein 
continuum beginning at $m=0$, where $k^2=-m^2$. When we restrict the 
coordinates
to the location of the $p$-brane at $U=U'=U_*$ we obtain the expression
(\ref{gravgf}).

When $p=5$ the solutions to the homogeneous equation are no longer Bessel
functions (\ref{sol5}).
Imposing the same boundary conditions as for $p\neq 5$ gives the Green's 
function
\begin{equation}
     {\cal G}_k(U,U^\prime) = -\frac{g_{YM}^2}{N} 
        \frac{1+\gamma}{2\gamma k^2} \frac{1}{U U'} 
         \left(\frac{U_*}{U_>}\right)^{-\gamma} 
        \left[ (\gamma-1) \left(\frac{U_*}{U_<}\right)^{\gamma}
              +(\gamma+1) \left(\frac{U_*}{U_<}\right)^{-\gamma}\right]~,
\end{equation}
where $\gamma=\sqrt{1+\lambda k^2}$. When $U=U'=U_*$ we recover the 
expression (\ref{d5gf}).
Notice also that from the pole condition $k^2=0$ there is a massless mode. 
In addition
from the pole condition $\gamma =0$, there is also a branch cut singularity
beginning at $m=1/{\sqrt{\lambda}}$. Thus when $p=5$, the Kaluza-Klein continuum is separated
from the zero mode by a mass gap $1/{\sqrt{\lambda}}$.

The Green's function for the gauge field can be obtained by following a 
similar procedure to the minimally coupled scalar. In the case of $p\neq 5$ the
expression is
\begin{eqnarray}
     {\cal G}_k(U,U^\prime) &=&i \frac{\pi g_{YM}^2 }{N(5-p)} \frac{1}{ UU'}
       \frac{H_{\alpha-1}^{(1)}(q_<)}{H_{\alpha-2}^{(1)}(q_*)} 
        \nonumber \\
       &&\qquad\qquad \times\left[J_{\alpha-2}(q_*)H_{\alpha-1}^{(1)}(q_>)
     - H_{\alpha-2}^{(1)}(q_*) J_{\alpha-1}(q_>) \right]~,
\end{eqnarray}
while for $p=5$ it is simply
\begin{equation}
     {\cal G}_k(U,U^\prime) = -\frac{1}{N^2} 
        \frac{1+\gamma}{2\gamma k^2} \frac{1}{U U'} 
       \left(\frac{U_*}{U_>}\right)^{-\gamma} 
         \left[ (\gamma-1) \left(\frac{U_*}{U_<}\right)^{\gamma}
              +(\gamma+1) \left(\frac{U_*}{U_<}\right)^{-\gamma}\right]~.
\end{equation}
Again we see that the $p+1$-dimensional Kaluza-Klein spectrum has the same
characteristics as that found for the minimally coupled scalar. 
When $U=U'=U_*$
we obtain the results (\ref{gaugegf}) for $p\neq5$, and 
(\ref{gfgf5}) for $p=5$.


\newpage


\begin{thebibliography}{99}

\bibitem{malda}
J.~Maldacena,
``{\it The large N limit of superconformal field theories and supergravity},''
Adv.\ Theor.\ Math.\ Phys.\ {\bf 2} (1998) 231
[hep-th/9711200].

\bibitem{Gubser:1998bc}
S.~S.~Gubser, I.~R.~Klebanov and A.~M.~Polyakov,
``{\it Gauge theory correlators from non-critical string theory},''
Phys.\ Lett.\ B {\bf 428}, 105 (1998)
[hep-th/9802109].

\bibitem{Witten:1998qj}
E.~Witten,
``{\it Anti-de Sitter space and holography},''
Adv.\ Theor.\ Math.\ Phys.\  {\bf 2}, 253 (1998)
[hep-th/9802150].

\bibitem{agmoo}
O.~Aharony, S.~S.~Gubser, J.~Maldacena, H.~Ooguri and Y.~Oz,
``{\it Large N field theories, string theory and gravity},''
Phys.\ Rept.\ {\bf 323} (2000) 183
[hep-th/9905111].

\bibitem{imsy}
N.~Itzhaki, J.~M.~Maldacena, J.~Sonnenschein and S.~Yankielowicz,
``{\it Supergravity and the large N limit of theories with sixteen  supercharges},''
Phys.\ Rev.\ {\bf D 58} (1998) 046004
[hep-th/9802042].

\bibitem{rsII}
L.~Randall and R.~Sundrum,
``{\it An alternative to compactification},''
Phys.\ Rev.\ Lett.\ {\bf 83} (1999) 4690
[hep-th/9906064].

\bibitem{Verlinde:2000fy}
H.~Verlinde,
``{\it Holography and compactification},''
Nucl.\ Phys.\ B {\bf 580}, 264 (2000)
[hep-th/9906182].

\bibitem{aio}
M.~Alishahiha, H.~Ita and Y.~Oz,
``{\it Graviton scattering on D6 branes with B fields},''
JHEP{\bf 0006} (2000) 002
[hep-th/0004011].

\bibitem{hash}
A.~Hashimoto and N.~Itzhaki,
``{\it A comment on the Zamolodchikov c-function and the black string 
entropy},''
Phys.\ Lett.\ B {\bf 454} (1999) 235
[hep-th/9903067].

\bibitem{boz}
A.~Brandhuber and Y.~Oz,
``{\it The D4-D8 brane system and five dimensional fixed points},''
Phys.\ Lett.\ B {\bf 460} (1999) 307
[hep-th/9905148].

\bibitem{youm}
D.~Youm,
``{\it Solitons in brane worlds},''
Nucl.\ Phys.\ B {\bf 576} (2000) 106
[hep-th/9911218];
D.~Youm,
``{\it Solitons in brane worlds II},''
Nucl.\ Phys.\ B {\bf 576} (2000) 139
[hep-th/0001018];
D.~Youm,
``{\it Bulk fields in dilatonic and self-tuning flat domain walls}
''
Nucl.\ Phys.\ B {\bf 589} (2000) 315
[hep-th/0002147];
D.~Youm,
``{\it Localization of Gravity on Dilatonic Domain Walls: 
Addendum to ``Solitons in Brane Worlds''},''
Nucl.\ Phys.\ B {\bf 596} (2001) 289
[hep-th/0007252].

\bibitem{clp}
M.~Cvetic, H.~Lu and C.~N.~Pope,
``{\it Domain walls with localised gravity and domain-wall/QFT 
correspondence},''
Phys.\ Rev.\ D {\bf 63} (2001) 086004
[hep-th/0007209].


\bibitem{Boonstra:1999mp}
H.~J.~Boonstra, K.~Skenderis and P.~K.~Townsend,
``{\it The domain wall/QFT correspondence},''
JHEP {\bf 9901}, 003 (1999)
[hep-th/9807137].


\bibitem{Behrndt:1999mk}
K.~Behrndt, E.~Bergshoeff, R.~Halbersma and J.~P.~van der Schaar,
``{\it On domain-wall/QFT dualities in various dimensions},''
Class.\ Quant.\ Grav.\  {\bf 16}, 3517 (1999)
[hep-th/9907006].


\bibitem{gubser}
J.~Maldacena, unpublished; E.~Witten, unpublished; 
S.~S.~Gubser,
``{\it AdS/CFT and gravity},''
Phys.\ Rev.\ D {\bf 63} (2001) 084017
[hep-th/9912001].

\bibitem{pomarol}
A.~Pomarol,
``{\it Gauge bosons in a five-dimensional theory with localized gravity},''
Phys.\ Lett.\ B {\bf 486} (2000) 153
[hep-ph/9911294].

\bibitem{kss}
N.~Kaloper, E.~Silverstein and L.~Susskind,
``{\it Gauge symmetry and localized gravity in M theory},''
JHEP {\bf 0105} (2001) 031
[hep-th/0006192].

\bibitem{apr}
N.~Arkani-Hamed, M.~Porrati and L.~Randall,
``{\it Holography and phenomenology},''
hep-th/0012148; 

\bibitem{rz}
R.~Rattazzi and A.~Zaffaroni,
``{\it Comments on the holographic picture of the Randall-Sundrum model},''
JHEP {\bf 0104} (2001) 021
[hep-th/0012248].

\bibitem{gs}
T.~Gherghetta and M.~Shaposhnikov,
``{\it Localizing gravity on a string-like defect in six dimensions},''
Phys.\ Rev.\ Lett.\ {\bf 85} (2000) 240
[hep-th/0004014].

\bibitem{poppitz}
E.~Ponton and E.~Poppitz,
``{\it Gravity localization on string-like defects in codimension two and the  AdS/CFT correspondence},''
JHEP{\bf 0102} (2001) 042
[hep-th/0012033].

\bibitem{grs}
T.~Gherghetta, E.~Roessl and M.~Shaposhnikov,
``{\it Living inside a hedgehog: Higher-dimensional solutions that localize  
gravity},'' Phys.\ Lett.\ B {\bf 491} (2000) 353
[hep-th/0006251];
T.~Gherghetta,
``{\it Localizing gravity on a 3-brane in higher dimensions},''
hep-th/0010276.

\bibitem{gkr}
S.~B.~Giddings, E.~Katz and L.~Randall,
``{\it Linearized gravity in brane backgrounds},''
JHEP {\bf 0003} (2000) 023
[hep-th/0002091].










\end{thebibliography}
\end{document}